\documentclass[a4paper]{jpconf}
\pdfoutput=1
\usepackage{mathletters}

\usepackage[dvipsnames]{xcolor}

\usepackage{bm}
\usepackage{amsmath}
\usepackage{amssymb}
\usepackage{graphicx}
\usepackage{hyperref}
\usepackage{cleveref}
\usepackage{slashed}
\usepackage{amsfonts}
\usepackage{amsbsy}
\usepackage{color}
\usepackage{array}
\usepackage{dcolumn}
\usepackage{tensor}
\usepackage{braket}
\usepackage{mathrsfs}
\usepackage{tikz}
\usepackage{verbatim}

\begin{document}
\title{From theory to precision modelling of strong-field QED in the transition regime}

\author{Alexander J. Macleod, on behalf of the LUXE collaboration}

\address{Institute of Physics of the ASCR, ELI-Beamlines, Na Slovance 2, 18221 Prague, Czechia}

\ead{alexander.macleod@eli-beams.eu}

\begin{abstract}
    The combination of energetic electron beams, delivered from conventional accelerators at a high repetition rate, and ultraintense lasers, makes it possible to perform precision measurements of strong-field QED. 
    The LUXE collaboration aims to perform precision measurements of nonlinear Compton scattering and Breit-Wheeler pair creation in the transition from the perturbative to nonperturbative regimes.  
    Here we present an overview of recent developments in the modelling of strong-field QED processes, which are needed to reach the required precision of a few percent for intensity parameters $0.1 < \xi < 10$. 
    We discuss how to go from plane-wave QED results to numerical simulations and present predicted signals and error estimates.
\end{abstract}

%%%%%%%%%%%%%%%%%%%%
\section{Introduction \label{sec:Intro}}
%%%%%%%%%%%%%%%%%%%%

Precision tests of QED, such as the measurement of the electron anomalous magnetic moment, have demonstrated extraordinary levels of agreement between theory and experiment.
This relies on the ability to perform calculations \emph{perturbatively} in the coupling\footnote{We use units where $\hbar = c = \epsilon_{0} = 1$ throughout.}, $\alpha = e^{2}/4\pi \simeq 1/137$.
Whether this agreement persists in the \emph{non-perturbative} regime of QED is an open question.

At high energies and background field strengths we enter the strong-field regime of QED, which requires a non-perturbative treatment of the interaction with the background field.
This regime is characterised by two key parameters becoming larger than unity.
Firstly, the intensity parameter, $\xi = e E_{0} \lambdabar_{\text{C}}/\omega_{0}$, which is a dimensionless measure of the work done on an electron by a laser of peak field strength $E_{0}$ and frequency $\omega_{0}$ over a (reduced) Compton wavelength $\lambdabar_{\text{C}}$~\cite{Ritus:1985,heinzl.oc.2009}. 
This is often called the ``classical nonlinearity parameter'', and describes the effective coupling to the background.
Secondly, the ``quantum nonlinearity parameter'' of a probe particle with momentum $p_{\mu}$, $\chi = e \sqrt{- (p \cdot F)^{2}}/m^{3}$, with $F_{\mu\nu}$ the background field tensor.
This gives a measure of the importance of quantum effects and recoil in the interaction.
In a plane wave background $\chi = \eta \xi$, where $\eta = k \cdot p/m^{2}$ is often called the ``lightfront energy'' of a probe particle with momentum $p_{\mu}$, where $k_{\mu}$ is the laser wavevector.

The intensity parameter, $\xi$, roughly corresponds to the number of background photons absorbed over a Compton wavelength, and so determines the onset of nonlinear effects.
When $\xi \ll 1$, processes are well approximated using a perturbative treatment.
However, when $\xi \gtrsim 1$ a fully \emph{non-perturbative} treatment of the interaction with the background is required. 
This poses several challenges for theory, particularly in developing analytical models which are useful for the analysis of laser-matter interactions.

\begin{figure*}[t!]
    \centering
    \includegraphics[width=0.55\linewidth]{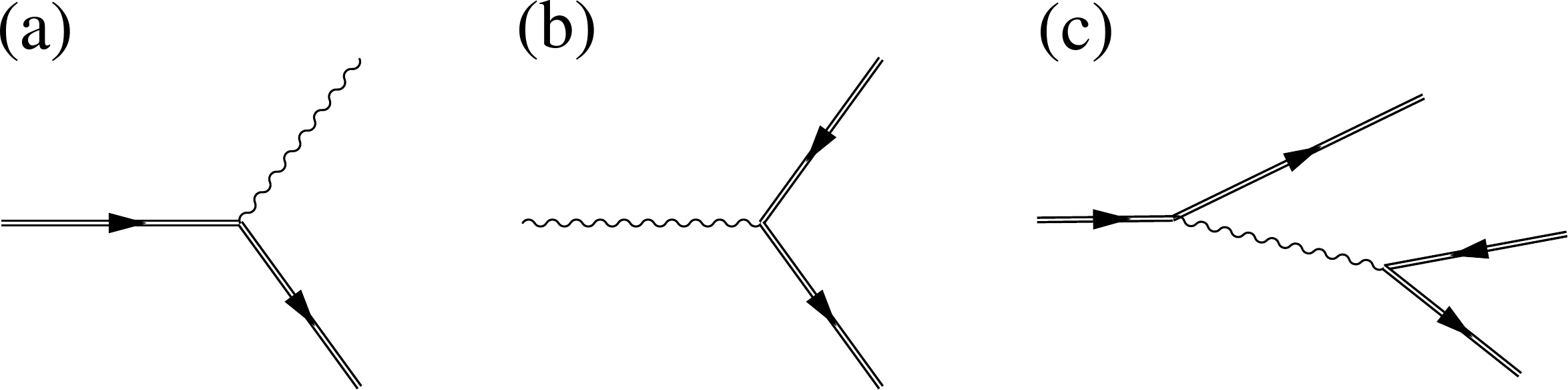}
    \caption{%
        The three key processes which will be measured at LUXE:
        (a) Nonlinear Compton scattering (b) Breit-Wheeler pair production (c) trident pair production.
    }
    \label{fig:Processes}
\end{figure*}

\begin{figure*}[b]
    \centering
    \includegraphics[width=0.85\linewidth]{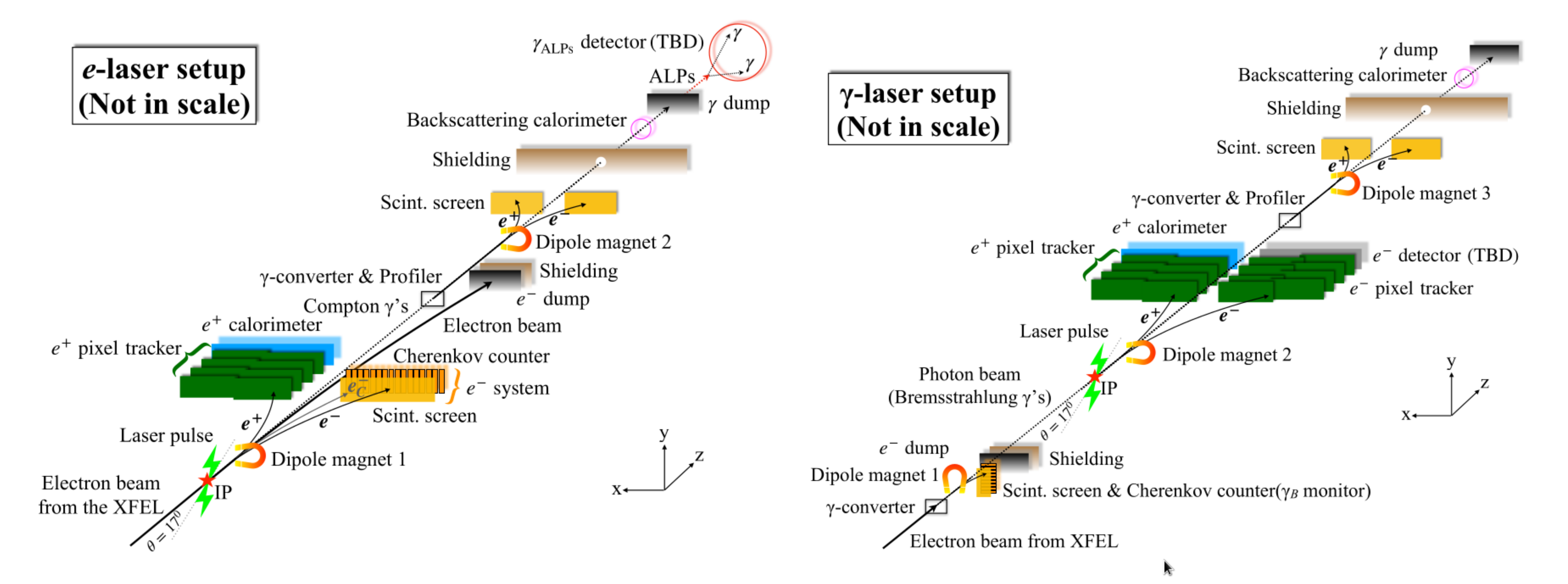}
    \caption{%
        Schematic experimental layouts for: (left) $e$-laser and (right) $\gamma$-laser setup~\cite{abramowicz2021conceptual}.
    }
    \label{fig:SchematicDetailed}
\end{figure*}

One such model, valid when $\xi \gg 1$ \emph{and} $\xi \gg \chi^{1/3}$, is the locally constant field approximation (LCFA)~\cite{Ritus:1985}.
It forms the basis of many simulation frameworks used to analyse QED effects in laser-matter interactions (e.g.~\cite{ridgers2014modelling,gonoskov2015extended,lobet2016modeling}). 
Local \emph{rates}, which encode quantum effects in these simulations, depend only on $\chi$ in the LCFA.
However, limitations of the LCFA~\cite{harvey2015testing,blackburn2018benchmarking,di2018implementing,ilderton2019extended,king2020uniform} in the \emph{transition regime} from perturbative to non-perturbative physics, where $\xi \sim \mcO(1)$, cause large errors when compared with exact QED results.
This problem is of immediate concern; a new wave of experiments exploring strong-field QED effects when $\xi \sim \mcO(1)$ are on the horizon.

\begin{figure*}[t!]
    \centering
    \includegraphics[width=0.45\linewidth]{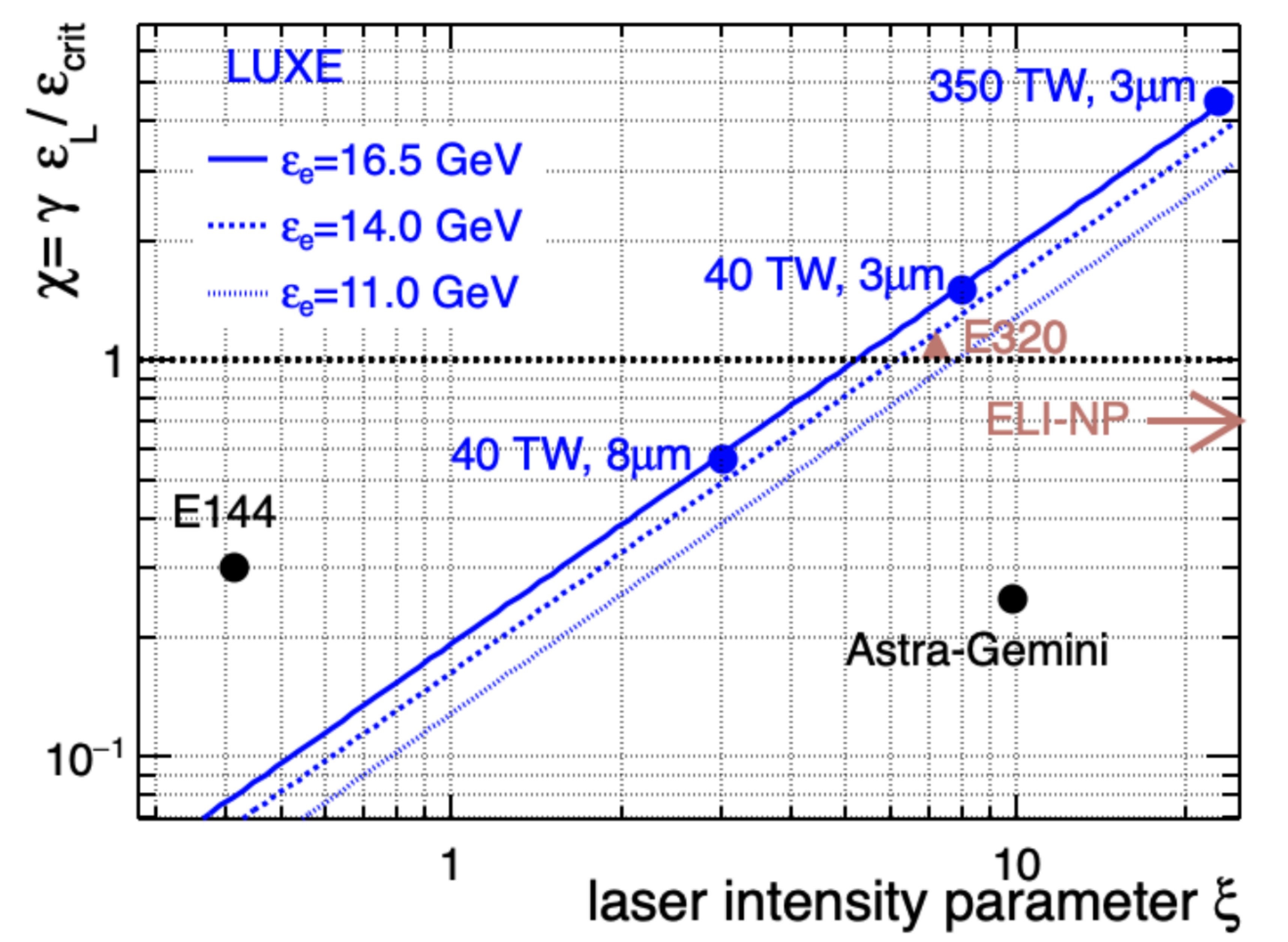}
    \caption{%
        Strong-field QED $\chi$ vs $\xi$ parameter space.
        Isolines represent three different electron beam energies $\varepsilon_{e}$.
        Blue dots show parameters accessible to LUXE in each phase~\cite{abramowicz2021conceptual}.
        Also shown are parameters for: E-144~\cite{bamber1999studies}, E-320~\cite{meuren2019probing} Astra-Gemini~\cite{cole2018experimental,poder2018experimental} and ELI-NP~\cite{gales2018extreme}.
    }
    \label{fig:Parameter}
\end{figure*}

The Light Und XFEL Experiment (LUXE)~\cite{abramowicz2021conceptual} will perform high-precision measurements of strong-field QED processes in the transition regime.
This will be achieved using a 16.5~GeV electron beam from the European XFEL and 40~TW JETI laser (Phase-0), with a subsequent upgrade to 350~TW (Phase-1), to measure nonlinear Compton scattering~\cite{Nikishov:1964zza,Brown:1964zzb} and trident pair production~\cite{Ritus:1972nf} in electron seeded interactions, and Breit-Wheeler pair production~\cite{Breit:1934zz,Reiss:1962,narozhny1969propagation} in photon seeded interactions (see~\cref{fig:Processes}).
Schematics for both the electron and photon seeded interactions are given in \cref{fig:SchematicDetailed}.
The parameters accessible to LUXE are shown in \cref{fig:Parameter}

To enable accurate modelling of the experimental outcomes of LUXE, and analysis of the experimental data, a robust simulation framework which works in the transition regime is required.
The \emph{locally-monochromatic approximation} (LMA) has been shown to agree with exact QED results~\cite{heinzl.pra.2020,Blackburn:2021rqm,blackburn2021higher} and can be used to derive local rates that take into account interference effects at the scale of the laser wavelength.
These rates have been implemented into the simulation framework PTARMIGAN, which describes strong-field QED effects through combining classical trajectories with probability rates that treat the background `locally' as a monochromatic plane wave.

The outline of this paper is as follows. 
In \cref{sec:Theory} we discuss how strong-field effects are modelled in the transition regime using the LMA, in particular outlining the steps from theory to the simulation framework PTARMIGAN.
The calculated experimental signatures of the transition regime for LUXE are presented in \cref{sec:Results}.
Finally, we conclude and summarise in \cref{sec:Summary}.

%%%%%%%%%%%%%%%%%%%%
\section{From theory to simulation in the transition regime \label{sec:Theory}}
%%%%%%%%%%%%%%%%%%%%

%%%%%%%%%%%%%%%
\subsection{Plane waves and interference \label{sec:PW}}
%%%%%%%%%%%%%%%

We begin with the Dirac equation for an electron, mass $m$ and charge $e$, coupled to a gauge potential describing the background field, $A_{\mu}$,
\begin{align}\label{eqn:Dirac}
    (i \slashed{\partial} - e \slashed{A} - m)
    \psi(x)
    =
    0
    \,.
\end{align}
Here, $\slashed{b} \equiv b_{\mu} \gamma^{\mu}$, with $\gamma^{\mu}$ the Dirac matrices.
To treat the background field non-perturbatively, we seek field configurations for which \cref{eqn:Dirac} can be solved \emph{exactly}.
This gives \emph{dressed} wavefunctions, $\psi(x)$, where the interaction with the background is included to all orders, and allows one to calculate scattering processes in the \emph{Furry picture}~\cite{Furry:1951zz}.

Exact solutions to the Dirac equation for focused laser pulses are not known.
However, if the initial Lorentz factor of the probe particle $\gamma \gg \xi^{2}$, the probability for a strong-field QED process in a focused background, $\sfP_{\text{focus}}$, can be expressed as a sum over the probability for the process to occur in a plane wave, $\sfP_{\text{pw}}$, weighted by the electron areal probe density, $\rho(x^{\lcperp})$~\cite{di2014ultrarelativistic},
\begin{align}\label{eqn:FocusApprox}
    \sfP_{\text{focus}}
    \approx
    \int \ud^{2} x^{\lcperp}
    \rho(x^{\lcperp})
    \sfP_{\text{pw}}\big[
        \xi(x^{\lcperp}),\eta(x^{\lcperp})
    \big] 
    \,,
\end{align}
where $x^{\lcperp}$ represent the directions transverse to the laser propagation direction and the intensity parameter, $\xi$, and lightfront energy, $\eta$, are defined \emph{locally}.
At LUXE, the electron beam energy of $\varepsilon_{e} \le 16.5$~GeV at the proposed collision angle of $\vartheta = 17.2^{\circ}$ corresponds to $\gamma \simeq 10^{4}$, and so to leading order \cref{eqn:FocusApprox} is a good approximation.
This motivates us to use the plane wave approximation to calculate the probabilities, and subsequently \emph{rates}, which will be used to simulate the experimental outcomes of LUXE.

Working in a plane wave background, it is convenient to use lightfront variables.
A plane wave propagating in the $z$-direction with light-like ($k^{2} = 0$) wavevector $k_{\mu} = \delta_{\mu}^{\lcm} k_{\lcm} = \omega (1,0,0,1)$ has a phase $\varphi \equiv k \cdot x = \omega x^{\lcm}$, where $x^{\pm} = t \pm z$.
Here, $x^{\lcm}$ acts as a ``time'' coordinate, with $x^{\lcp}$ the \emph{longitudinal} direction, and the remaining coordinates $x^{\lcperp} = (x,y)$ being \emph{transverse} to the propagation direction.
Solving \cref{eqn:Dirac} in the gauge potential $e A_{\mu}(\varphi) = a_{\mu}(\varphi)$, one arrives at the \emph{Volkov solutions},
\begin{align}\label{eqn:Volkov}
    \psi_{p}(x)
    =
    \bigg(
        1 + \frac{\slashed{k} \slashed{a}(\varphi)}{2 k \cdot p}    
    \bigg)
    u_{p}
    \,
    e^{- i S_{\text{cl}}(x)}
    \,,
\end{align}
for an electron with initial momentum $p_{\mu}$ and constant spinor $u_{p}$.
The exponent, $S_{\text{cl}}(x)$, is the \emph{classical action} for an electron in a plane wave field with instantaneous momentum, $\pi_{\mu}$,
\begin{align}\label{eqn:Momentum}
    S_{\text{cl}}(x)
    =
    p \cdot x
    + 
    \int^{\varphi}
    \!
    \ud t
    \,
    \frac{2 p \cdot a(t) - a^{2}(t)}{2 k \cdot p}
    \,,
    \quad
    \pi_{\mu}(\varphi)
    =
    p_{\mu}
    -
    a_{\mu}(\varphi)
    +
    \frac{2 p \cdot a(\varphi) - a^{2}(\varphi)}{2 k \cdot p}
    k_{\mu}
    \,.
\end{align}

On a plane wave background only three components of energy and momentum are conserved, and the $S$-matrix for a given process can always be expressed as,
\begin{align}\label{eqn:SMatrix}
    S_{\text{fi}}
    =
    (2\pi)^{3}
    \delta_{\lcp,\lcperp}^{3}(P_{\text{in}} - P_{\text{out}})
    \mcM
    \,,
\end{align}
where $\mcM$ is the invariant part of the amplitude, $P_{\text{in}}$ and $P_{\text{out}}$ are the sums of the incoming and outgoing momenta, respectively, and $\delta_{\lcp,\lcperp}^{3}(q) \equiv \delta(q_{\lcp})\delta(q_{x})\delta(q_{y})$.
The non-conservation of all components of the momentum gives rise to the possibility of processes which are kinematically forbidden in vacuum, such as the target processes of LUXE (see~\cref{fig:Processes}).

Consider a first-order processes, such as non-linear Compton scattering or Breit-Wheeler pair production (see \cref{fig:Processes}(a) \& (b)).
The probability takes the form $\sfP_{\text{pw}} \sim \int \ud \Omega_{\text{LIPS}} \int \ud \varphi \ud \varphi^{\prime} |\mcL_{\text{I}}|^{2}$, where $\mcL_{\text{I}}$ is the interaction Lagrangian and $\ud \Omega_{\text{LIPS}}$ is the Lorentz invariant phase space for the process.
This integral is non-local; interference between different spacetime points in the field are of crucial importance.
This non-locality makes the probability unsuitable for numerical simulations, which require \emph{local rates}.
Defining the average $\phi = (\varphi + \varphi^{\prime})/2$ and interference $\theta = \varphi - \varphi^{\prime}$ phases, one method of obtaining a local description of the probability is to expand the probability as a series in $\theta \ll 1$.
This is the LCFA: interference effects are then completely neglected which leads to large errors in the transition regime~\cite{harvey2015testing,blackburn2018benchmarking,di2018implementing,ilderton2019extended,king2020uniform}.
The experimental outcomes of LUXE are analysed using a different approach.

%%%%%%%%%%%%%%%
\subsection{Locally-monochromatic approximation \label{sec:LMA}}
%%%%%%%%%%%%%%%

Instead of performing an expansion directly in the interference phase, $\theta$, one can include interference effects at the scale of the laser wavelength using the \emph{locally-monochromatic approximation} (LMA).
A typical plane wave pulse will have both slow and fast oscillations.
Consider a circularly polarised plane wave pulse,
\begin{align}\label{eqn:Pulse}
    a_{\mu}(\varphi)
    =
    m \xi_{0} f\Big(\frac{\varphi}{\Phi}\Big)
    \big[
        \alpha_{\mu} \cos\big(b(\varphi)\big)
        +
        \beta_{\mu} \sin\big(b(\varphi)\big)
    \big] 
    \,,
\end{align}
where $\alpha_{\mu},\beta_{\mu}$ are spacelike vectors satisfying $\alpha^{2} = \beta^{2} = -1$ and $\alpha \cdot \beta = \alpha \cdot k = \beta \cdot k = 0$, $b(\varphi)$ accounts for chirp effects and $\xi_{0}$ is the intensity parameter.
The pulse envelope $f(\varphi/\Phi)$ is slowly-varying over the pulse phase duration, $\Phi$, while the carrier frequency terms in square brackets are rapidly oscillating.
In the LMA, the fast oscillations are included \emph{exactly}, and only the leading order contribution from a derivative expansion of the pulse envelope is taken into account.
This is essentially a combination of two approximations, the \emph{slowly-varying envelope approximation}~\cite{narozhny1996photon,mcdonald1997relativistic,seipt2016analytical} and a local expansion akin to that used in the LCFA~\cite{Ritus:1985,harvey2015testing,di2018implementing,ilderton2019extended,Nikishov:1964zza}.

This has both advantages and disadvantages over the LCFA.
Interference effects are included, the low-energy limit is exact~\cite{heinzl.pra.2020}, and physics in the transition regime is more accurately described.
However, it is only suitable for fields which can be well approximated by plane waves, and so cannot be used in environments with significant plasma production.
There is an explicit trade-off between versatility and accuracy in the transition regime.
The experimental parameters for LUXE~\cite{abramowicz2021conceptual} make the LMA the ideal candidate for modelling strong-field QED processes, and as such it is the central approximation used in numerical codes.

Consider nonlinear Compton scattering in the circularly polarised background, \cref{eqn:Pulse}. 
An electron with initial momentum $p_{\mu}$, corresponding to a lightfront energy $\eta_{0} = k \cdot p/m^{2}$, collides with the background (wavevector $k_{\mu}$) and emits a photon of momentum $l_{\mu}$, which we use to define the lightfront fraction of the emitted photon, $u = k \cdot l/k \cdot p$.
The probability can be expressed as, $\sfP_{\text{LMA}} = \int \ud\tau W_{\text{LMA}}(\tau)$, where the \emph{local rate}, $W_{\text{LMA}}$, takes the form,
\begin{align}\label{eqn:LMAProb}
    W_{\text{LMA}}(\tau)
    =
    \sum_{n = 1}^{\infty}
    \int_{0}^{u_{n,*}(\tau)}
    \ud u
    \frac{\ud^{2} \sfP_{n}^{\text{mono}}[\xi(\tau),\eta(\tau)]}{\ud \tau \ud u}
    \,,
\end{align}
with $\tau$ the proper time, related to the phase by $\ud \tau / \ud \phi = 1/ (m \eta_{0})$.
This is expressed explicitly in terms of the probability, $\sfP_{n}^{\text{mono}}$, for nonlinear Compton scattering into the $n$th harmonic in a \emph{monochromatic background} (see e.g.~\cite{Ritus:1985}).
However, instead of depending only on the initial intensity parameter, $\xi_{0}$, one finds that the intensity takes a local value given by, 
\begin{align}\label{eqn:LocalXi}
    \xi^{2}(\tau)
    =
    -
    \frac{a^{2}[\phi(\tau)]}{m^{2}}
    =
    \xi_{0}^{2}
    f^{2}\Big(\frac{\phi(\tau)}{\Phi}\Big)
    \,,
\end{align}
which is the component of the potential that varies slowly with respect to wavelength.
Furthermore, both the lightfront energy, $\eta(\tau) = \eta_{0} \omega(\tau)$, where the non-constant frequency $\omega(\tau) = \ud b[\phi(\tau)]/\ud \phi$ encodes laser pulse chirping, and the harmonic range, $u_{n,*}(\tau) = u_{n}(\tau)/(1 + u_{n}(\tau))$ where $u_{n}(\tau) = 2 n \eta(\tau)/(1 + \xi^{2}(\tau))$ is the edge of the classical (nonlinear) harmonic range, are phase (or proper time) dependent.
Armed with the LMA rate for a given process, such as \cref{eqn:LMAProb} for nonlinear Compton scattering or analogous expressions for Breit-Wheeler pair production (see~\cite{heinzl.pra.2020,blackburn2021higher}), the next step is to implement this into a numerical simulation.

%%%%%%%%%%%%%%%
\subsection{Simulations in the transition regime with PTARMIGAN \label{sec:PTARMIGAN}}
%%%%%%%%%%%%%%%

Simulation frameworks which include strong-field QED effects (e.g.~\cite{ridgers2014modelling,gonoskov2015extended,bamber1999studies,chen1995cain}) typically combine classical particle trajectories with probability rates~\cite{blackburn2020radiation}.
The information encoded in the rates determines the dynamical quantities that must be obtained from the classical trajectories.

\begin{figure*}
    \centering
    \includegraphics[width=0.7\linewidth]{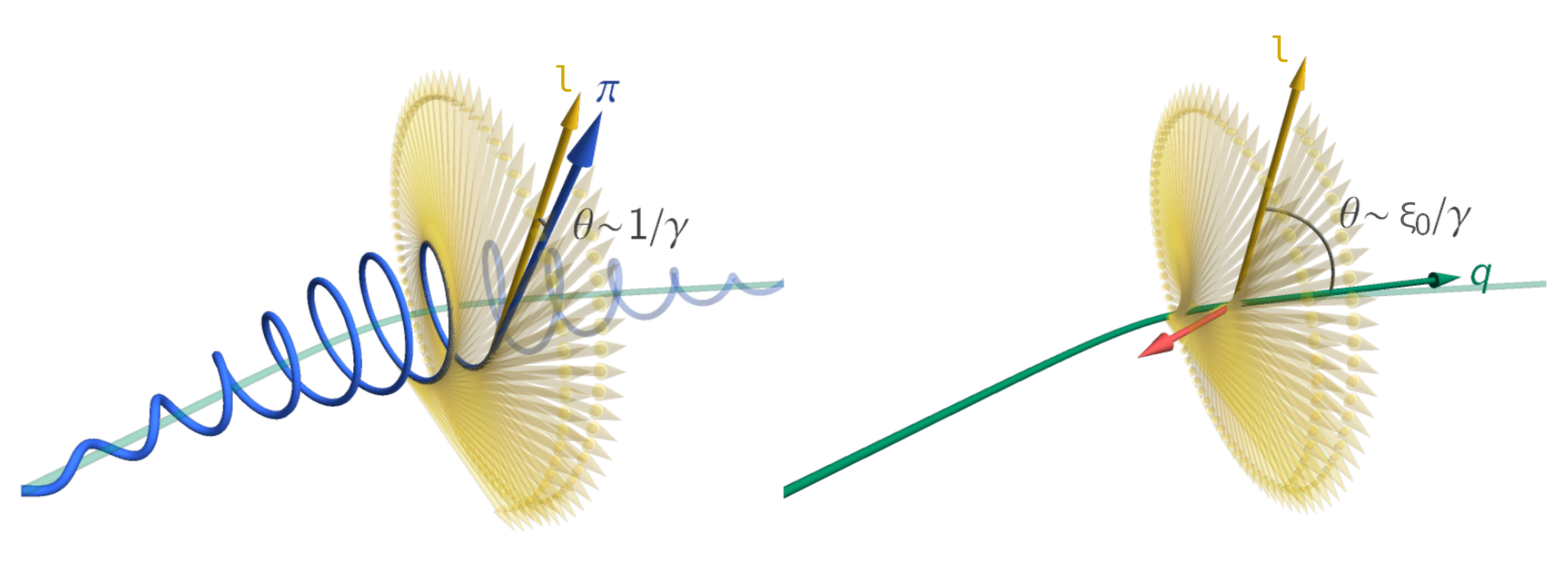}
    \caption{%
        Illustration of photon emission models: LCFA (left), LMA (right).
        Yellow arrow denotes the emitted photon, momentum $l_{\mu}$.
        Laser wavevector, $k_{\mu}$, shown in red.
        LCFA trajectory (blue) determined by the Lorentz force, with the momentum given by $\pi_{\mu}$ (c.f.~\cref{eqn:Momentum}).
        In the LMA, fast oscillations due to the carrier frequency are included exactly in the quantum rates, so the electron trajectory (green) is due to the ponderomotive force, with quasimomentum $q_{\mu} = \braket{\pi_{\mu}}$ the cycle average of $\pi_{\mu}$.
        Figure reproduced from Ref~\cite{Blackburn:2021rqm}.
    }
    \label{fig:Concept}
\end{figure*}

In the LCFA, one defines the classical trajectories of particles in terms of the kinetic momentum \cref{eqn:Momentum}.
Particles propagate between emission events due to the Lorentz force,
\begin{align}\label{eqn:Lorentz}
    \frac{\ud \pi_{\mu}}{\ud \tau}
    = 
    &
    - \frac{e}{m} F_{\mu\nu} \pi^{\mu}
    \,,
    \quad
    &
    \frac{\ud x^{\mu}}{\ud \tau}
    =
    &
    \frac{\pi^{\mu}}{m}
    \,,
\end{align}
with $x^{\mu}(\tau)$ the classical worldline.
This is an oscillatory trajectory, and emission occurs at a characteristic angle $\sim 1/\gamma$.
Contrastingly, LMA rates already treat oscillations due to the carrier frequency exactly.
The classical trajectory must reflect this, and upon averaging the motion over the fast timescale (the laser period) one finds that classical motion is determined by the relativistic ponderomotive force equation for the \emph{quasimomentum} $q_{\mu} = \braket{\pi_{\mu}}$, i.e. the laser-cycle-average of the instantaneous electron momentum in \cref{eqn:Momentum},
\begin{align}\label{eqn:Ponder}
    \frac{\ud q_{\mu}}{\ud \tau}
    =
    &
    \frac{m}{2}
    \partial_{\mu}
    \xi^{2}(\tau)
    \,,
    \quad
    &
    \frac{\ud x^{\mu}}{\ud \tau}
    =
    &
    \frac{q^{\mu}}{m \sqrt{1 + \xi^{2}(\tau)}}
    \,.
\end{align}
Emission now occurs in an angle $\sim \xi_{0}/\gamma$.
For an illustrative comparison of the particle trajectories in LCFA and LMA based simulations see \cref{fig:Concept}.

PTARMIGAN uses rates derived in the LMA combined with classical motion due to the relativistic ponderomotive force. 
It uses a Monte Carlo algorithm which:
\begin{enumerate}
    \item advances the electron trajectory by solving the ponderomotive force \cref{eqn:Ponder},

    \item evaluates at every time step the probability of emission and pseudorandomly decides whether to emit a photon or not, 

    \item where emission takes place, selects a harmonic index with probability $W_{n}/W$, where $W_{n}$ is the partial rate and $W = \sum_{n}^{\infty} W_{n}$ is the total rate,

    \item samples $u$, the lightfront momentum fraction, from the partial spectrum $(\ud W_{n}/\ud u)/W_{n}$,

    \item finally determines the emitted photon momentum, $k_{\mu}$, given $n$, $u$ and $q$, and resets the electron quasimomentum from $q$ to $q^{\prime}$.
\end{enumerate}
PTARMIGAN has been benchmarked for both nonlinear Compton scattering~\cite{Blackburn:2021rqm} and Breit-Wheeler pair production~\cite{blackburn2021higher}, showing excellent agreement to exact result in QED.
It is the fundamental tool underpinning analysis of the experimental outcomes at LUXE, and has been used to determine the non-perturbative signatures which will be observed.

%%%%%%%%%%%%%%%%%%%%
\section{Signatures of non-perturbative charge-field coupling at LUXE \label{sec:Results}}
%%%%%%%%%%%%%%%%%%%%

The flagship experimental campaigns at LUXE will measure non-perturbative charge-field coupling in nonlinear Compton scattering and Breit-Wheeler pair production.
PTARMIGAN has been used with realistic parameters for both the laser pulse and electron beam to analyse the non-perturbative signals.

%%%%%%%%%%%%%%%
\subsection{Non-perturbative signatures in nonlinear Compton scattering \label{sec:NLCNonpert}}
%%%%%%%%%%%%%%%

Nonlinear Compton scattering is the emission of a photon by an electron due to the interaction with a strong electromagnetic field, shown in \cref{fig:Processes}(a).
The process is distinguished from \emph{linear} Compton scattering, where only a single photon is absorbed from the background. 
The non-linear interaction with the background when $\xi \gtrsim 1$ causes higher order terms in a perturbative expansion in $\xi$ to dominate, requiring an all-orders, non-perturbative, treatment.

\begin{figure*}
    \centering
    \includegraphics[width=0.6\linewidth]{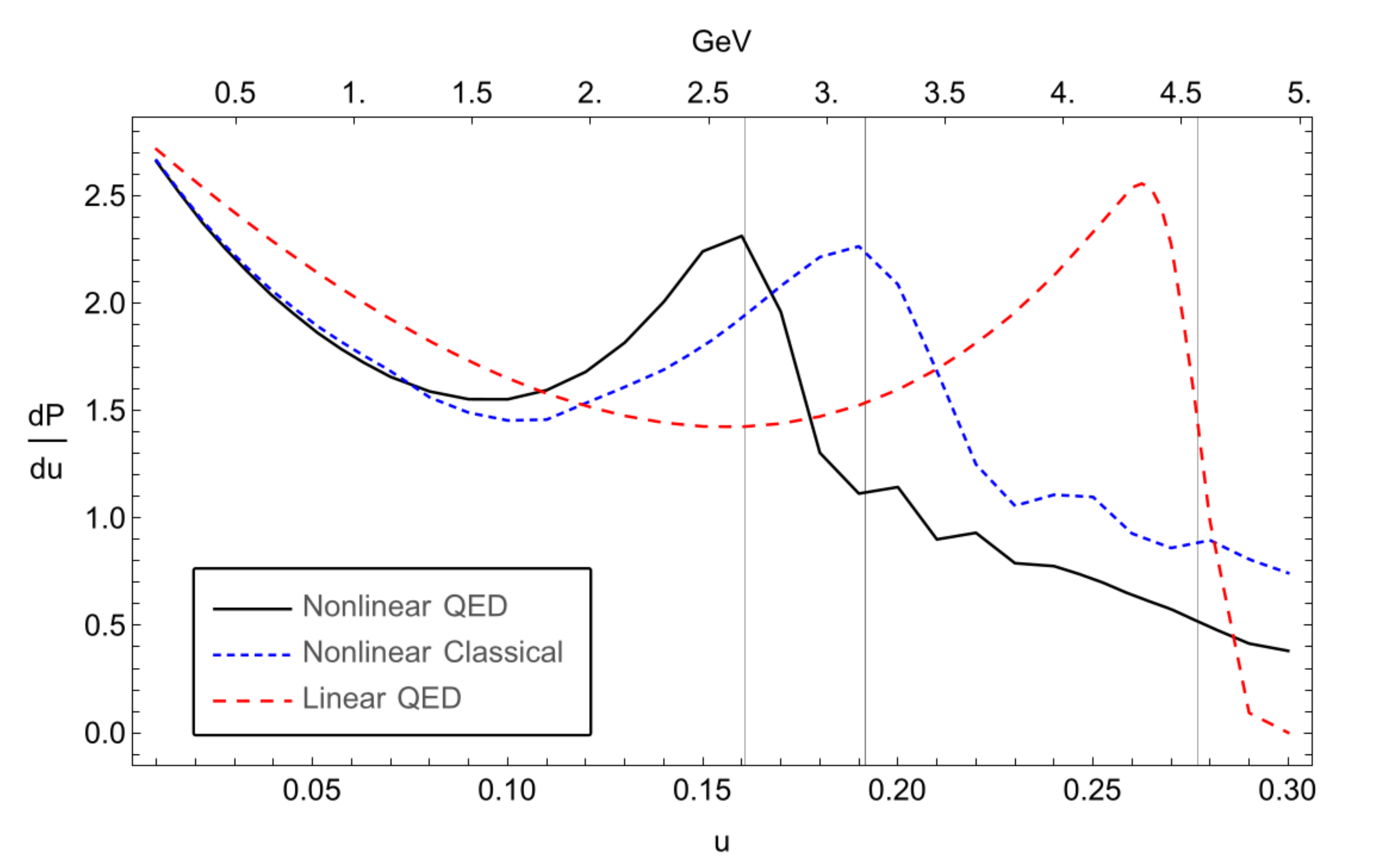}
    \caption{%
        Photon energy spectrum due to the interaction of an electron with energy $16.5$~GeV in different physical models: linear QED (red dashed), nonlinear classical (blue dotted), nonlinear QED (black solid)~\cite{abramowicz2021conceptual}.
        The position of the $n = 1$ harmonic peak, often called the Compton edge, is determined kinematically and plotted as a grey vertical line in each case.
    }
    \label{fig:NLCSig}
\end{figure*}

The key non-perturbative signature appears in the energy spectrum.
Shown in \cref{fig:NLCSig} is the photon energy spectrum versus $u = k \cdot l/k \cdot p$, for a $16.5$~GeV electron colliding with a plane wave laser pulse with intensity parameter $\xi_{0} = 1$ in: linear QED (red dashed), nonlinear classical electrodynamics (blue dotted) and nonlinear QED (black solid).
The first harmonic in the spectrum is clearly defined by the main peak.
This is the \emph{Compton edge}~\cite{harvey2009signatures}.

In linear QED, i.e. perturbative Compton scattering, the position of the Compton edge is $u_{\text{lin.QED}} = \frac{2 \eta_{0}}{1 + 2 \eta_{0}}$.
Multiple harmonic peaks become visible in the energy spectrum when nonlinear effects are included, with their position dependent on $\xi$.
In the nonlinear \emph{classical} theory, Thomson scattering, the position of the $n$th harmonic peak is $u^{n}_{\text{nonlin.class}} = \frac{2 n \eta_{0}}{1 + \xi^{2}}$.
Comparing the position of the Compton edge ($n = 1$) to the linear QED result, there is a clear red-shifting due to the nonlinear interaction.
Recoil effects are included in the transition from nonlinear classical electrodynamics to nonlinear QED leading to a further red-shifting of the harmonics to
$u^{n}_{\text{nonlin.QED}} = \frac{2 n \eta_{0}}{2 n \eta_{0} + 1 + \xi^{2}}$.
The red-shifting of the nonlinear QED Compton edge ($n = 1$) compared to linear QED is most readily explained by the electron gaining an \emph{effective mass}~\cite{harvey2009signatures}, which is due to the interaction with many background field photons.

\begin{figure*}
    \centering
    \includegraphics[width=0.6\linewidth]{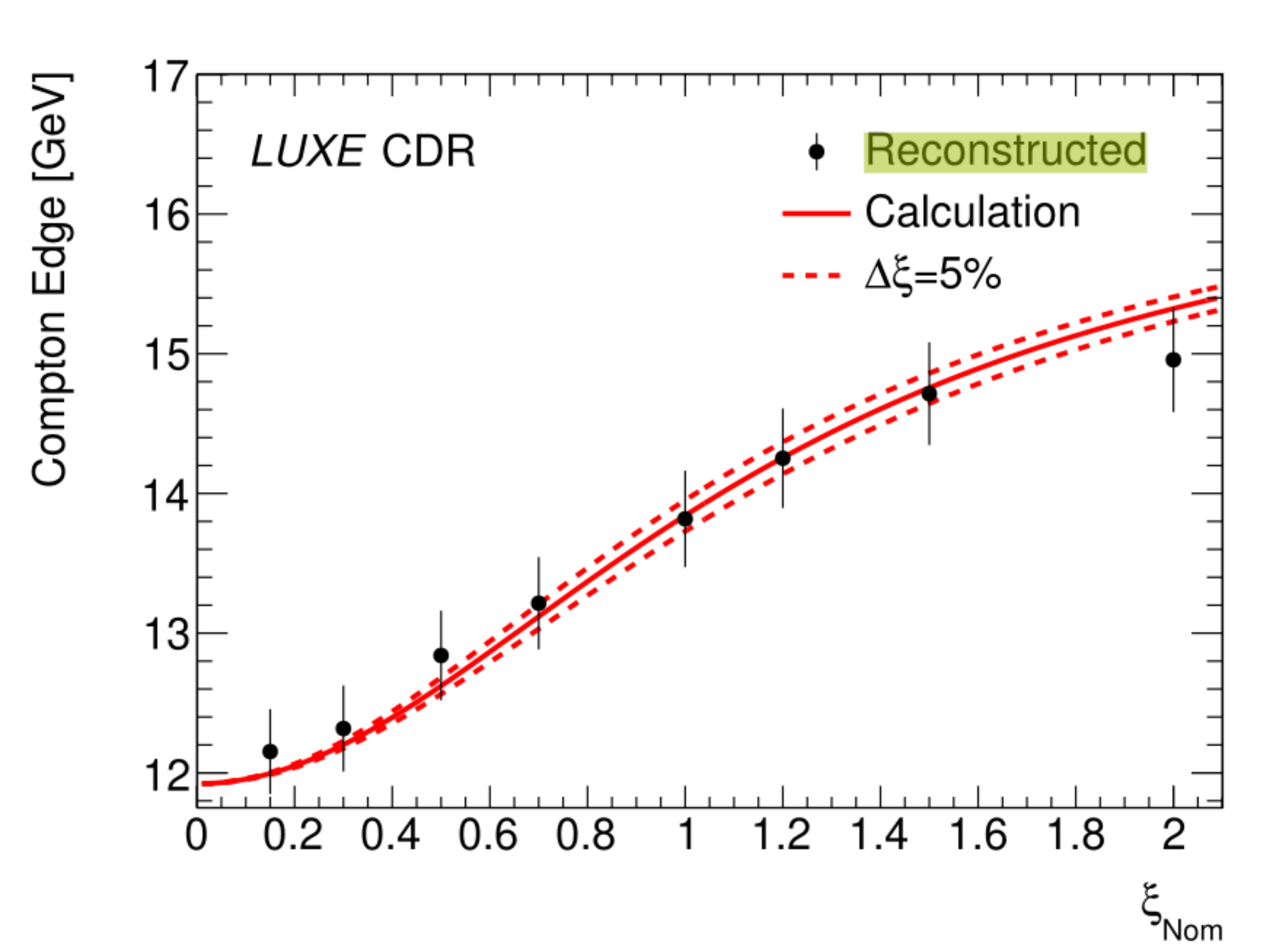}
    \caption{%
        Position of the leading Compton edge in the electron energy spectra as a function of the (nominal) intensity parameter, $\xi_{\text{nom}}$.
        Red solid line shows theoretical prediction, with dashed lines representing a $5$\% uncertainty in the intensity parameter value.
        Black data points calculated using PTARMIGAN for realistic beam and laser parameters (see~\cite{abramowicz2021conceptual} for details).
    }
    \label{fig:NLCRes2}
\end{figure*}

The shift of the leading Compton edge ($n = 1$) in the electron energy spectrum is shown in \cref{fig:NLCRes2}.
The electron beam has an initial energy of $16.5$~GeV, with a finite size of $\sigma_{\text{el}} = 5~\mu$m.
The laser pulse is taken to have a Gaussian profile in both the longitudinal and transverse directions.
The red solid line shows the theoretical prediction, with the dashed lines representing a $5\%$ uncertainty on the value of the laser intensity.
The black data points show the anticipated data result~\cite{abramowicz2021conceptual}, where the uncertainty in these data points is dominated by an energy scale uncertainty of $2.5\%$.
Here $\xi_{\text{nom}}$ is the \emph{nominal} value of the intensity parameter, which takes into account that the highest value $\xi$ is only in the peak of the pulse~\cite{abramowicz2021conceptual}.

%%%%%%%%%%%%%%%
\subsection{Non-perturbative signatures in Breit-Wheeler pair production \label{sec:BWNonpert}}
%%%%%%%%%%%%%%%

Breit-Wheeler pair creation is the decay of a photon into an electron positron pair due to the interaction with an intense electromagnetic field, see \cref{fig:Processes}(b).
As with nonlinear Compton scattering, the nonlinear interaction with the background field leads to an effective mass effect which requires an all-orders non-perturbative treatment of the background field.
This manifests as an increase in the threshold harmonic required for the photon to decay into an electron positron pair. 
Calculated in the LMA, this threshold is $n_{\star}(\phi) = \frac{2 (1 + \xi^{2}(\phi))}{\eta_{\gamma 0}}$, where $\xi(\phi)$ is defined analogously to \cref{eqn:LocalXi}, and $\eta_{\gamma 0}$ is the lightfront energy of the probe photon with momentum $l_{\mu}$, $\eta_{\gamma 0} = k \cdot l/m^{2}$ which we use to define the analogue of the quantum nonlinearity parameter for the photon, $\chi_{\gamma} = \eta_{\gamma 0} \xi$.
This dependence on the (local) intensity parameter signifies non-perturbativity at small coupling.

\begin{figure*}
    \centering
    \includegraphics[width=0.8\linewidth]{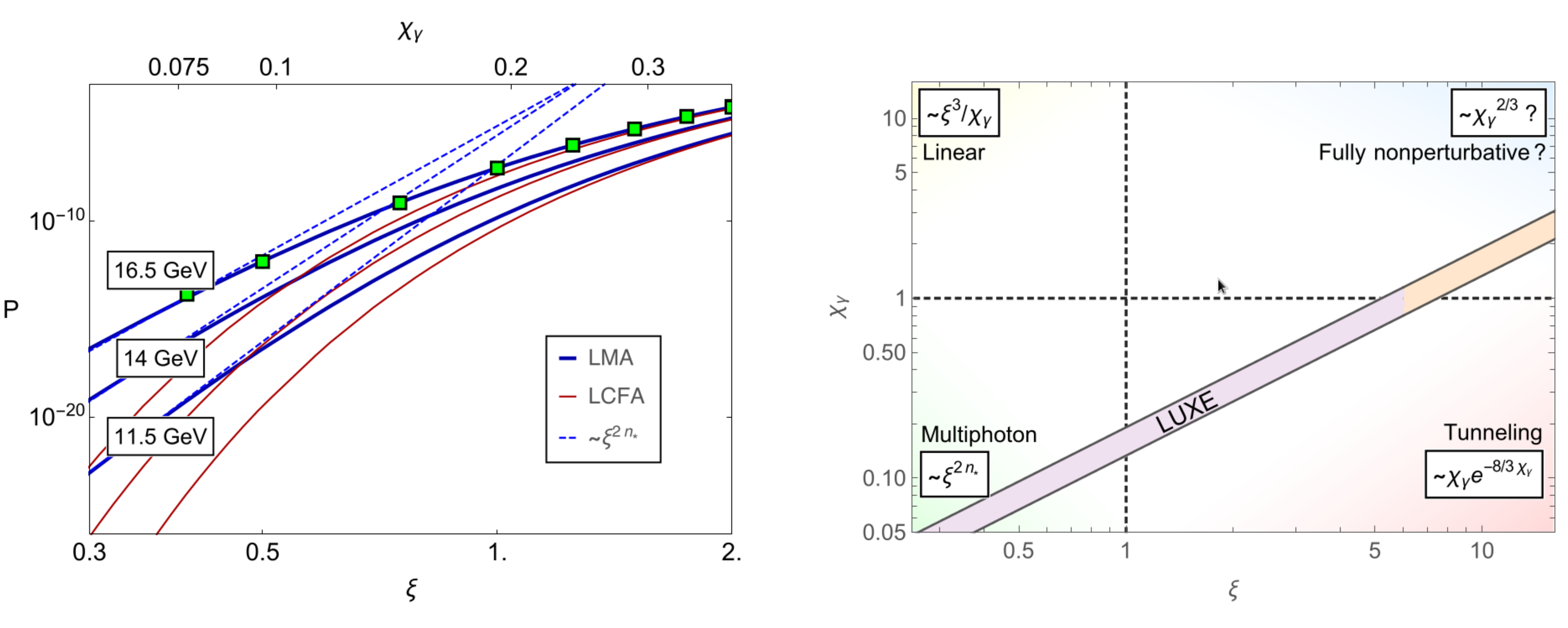}
    \caption{%
        (left) Probability of pair production due to the interaction of a high energy probe with different initial energies in the interaction with the laser pulse.
        Blue dashed lines correspond to pair production in the multiphoton, perturbative, regime.
        Green data points are the result of an exact analytical QED calculation using a plane wave background.
        Blue solid lines show LMA calculation, and excellent agreement is seen in the $16.5$~GeV case.
        Red lines show LCFA result, which becomes more accurate at high $\xi$, but fails at low values.
        (right) Pair production $\xi$, $\chi_{\gamma}$ parameter space explored by LUXE~\cite{abramowicz2021conceptual}.
    }
    \label{fig:BWSig}
\end{figure*}

For $\xi^{2} \ll 1$ and $\chi_{\gamma} \ll 1$ Breit-Wheeler pair production is perturbative, occurring as a ``multiphoton'' process in which the probability scales as $\sfP \sim \xi^{2 n_{\star}}$.
However, as the intensity parameter increases to $\xi \gtrsim 1$ with $\chi_{\gamma} \ll 1$, pair production occurs as a tunnelling-like process.
In the limit of a constant background, in this tunnelling regime the probability scales as $\sfP \sim \chi_{\gamma} \exp(- 8/3 \chi_{\gamma})$, and since $\chi_{\gamma}^{2} \propto \alpha$, this is non-perturbative in the charge field coupling.
This transition from the multiphoton into the tunnelling regime of Breit-Wheeler pair production will be experimentally measured at LUXE.
The parameter region accessible to LUXE is shown in the right hand plot of \cref{fig:BWSig}.

The key experimental signature is a characteristic ``turning of the curve'' in the probability (or pair-yield) vs $\xi$.
This is demonstrated in the left hand plot of \cref{fig:BWSig}, which shows the probability as a function of $\xi$ for different probe particle energies~\cite{abramowicz2021conceptual}.
The dashed lines correspond to the multiphoton process, when $\xi \ll 1$, while the blue solid lines show the results of simulation based on the LMA.
The ``turning of the curve'' refers to the pivoting of the probability away from the multiphoton result.
The data points show the analytical QED plane wave results for a photon energy of 16.5 GeV, which agree with the predictions of the LMA.
Also shown is the corresponding probability calculated using the LCFA in red solid lines, which clearly demonstrates the failure of the approach at low $\xi$.

\begin{figure*}
    \centering
    \includegraphics[width=0.8\linewidth]{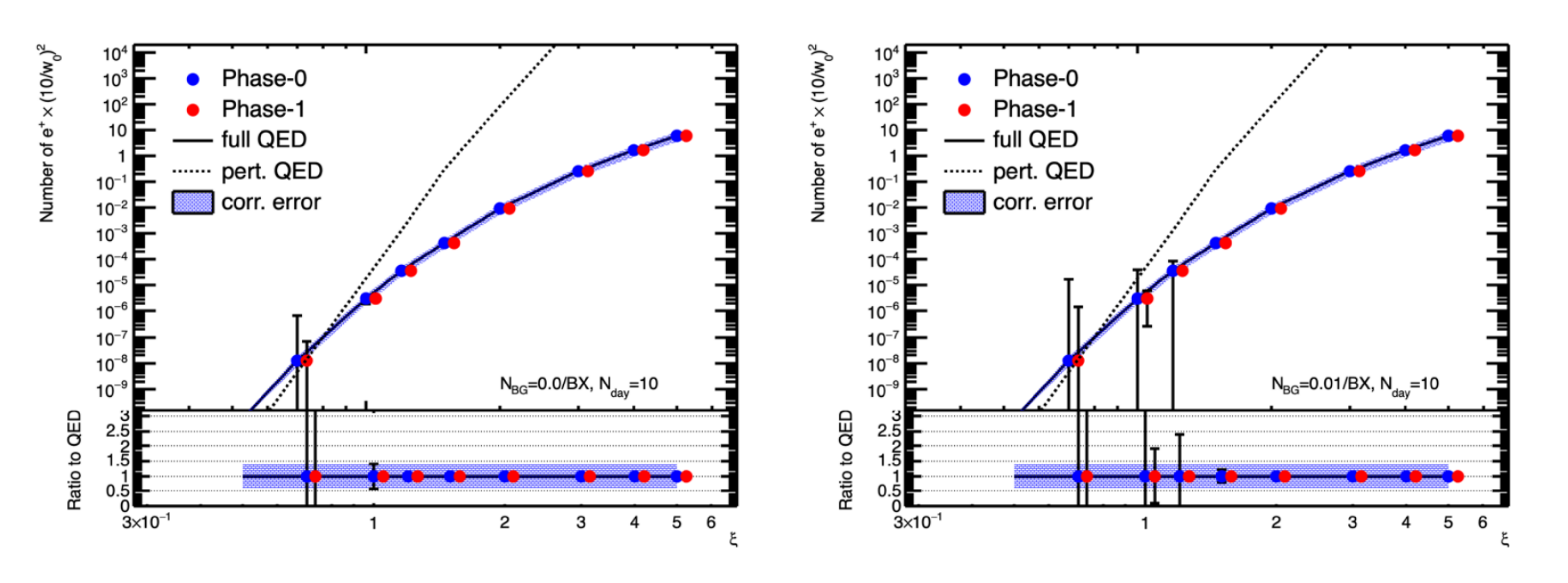}
    \caption{%
        Number of positrons per laser shot, normalised by $10/w_{0}^{2}$ versus $\xi$ for phase-0 and phase-1 at LUXE, with $w_{0}$ the focal spot size.
        The left (right) plot shows uncertainties corresponding to no particle background (a particle background of 0.01 particles per beam crossing).
        Dashed lines correspond to theoretical prediction of perturbative QED, solid lines show theoretical prediction of full strong-field QED.
        Bottom panels show ratio to QED prediction.
        More details in~\cite{abramowicz2021conceptual}.
    }
    \label{fig:BWRes}
\end{figure*}

Turning now to the predicted signal for the experimental parameters of LUXE, \cref{fig:BWRes} is the results from the simulation framework PTARMIGAN of the number of positrons produced per laser shot.
This shows the result one could obtain from 10 days of data taking per $\xi$ value (data points, blue: Phase-0, red: Phase-1), assuming no background particles per bunch crossing (left) and 0.01 background particles per bunch crossing (right).
The dotted lines show the results of perturbative QED, while the solid line shows the strong-field QED prediction in a plane wave pulse.
The errors at low $\xi$ are dominated by the relatively low number of pairs produced pairs, and these can be improved by suppression of the number of background particles.
For high intensity parameters, $\xi > 2$, statistical precision falls well below 5\%.

%%%%%%%%%%%%%%%%%%%%
\section{Summary \label{sec:Summary}}
%%%%%%%%%%%%%%%%%%%%

LUXE will explore the transition from perturbative to non-perturbative charge-field coupling.
To enable the experimental outcomes of LUXE to be analysed, a new simulation framework, PTARMIGAN, has been developed that models strong-field QED effects using the locally-monochromatic approximation.
This framework has been extensively benchmarked, and used to simulate the expected signals of non-perturbative physics which will be measured at LUXE.
These are the shifting of the Compton edge in the electron/photon energy spectra in nonlinear Compton scattering, and the ``turning of the curve'' signifying pair production via a tunnelling mechanism in the Breit-Wheeler process.

\subsection{Acknowledgments}

AJM is supported by the project High Field Initiative (CZ.02.1.01/0.0/0.0/15\_003/0000449) from the European Regional Development Fund.
This work has benefited from computing services provided by the German National Analysis Facility (NAF).

\section*{References}

\bibliographystyle{iopart-num}
\bibliography{Bib}

\end{document}